\documentclass[aps,prl,preprint,superscriptaddress]{revtex4}

\usepackage{graphicx}
\usepackage{dcolumn}
\usepackage{bm}

\begin{document}

\preprint{Los Alamos National Laboratory LA-UR-07-1934}

\title{Hidden Structure in Protein Energy Landscapes}



\author{Dengming Ming}
\affiliation{Computer, Computational, and Statistical Sciences Division}

\author{Marian Anghel}
\affiliation{Computer, Computational, and Statistical Sciences Division}

\author{Michael E. Wall}
\affiliation{Computer, Computational, and Statistical Sciences Division}
\affiliation{Bioscience Division and Center for Nonlinear Studies\\Los Alamos National Laboratory, Los Alamos, NM 87545, USA}
\email[E-mail: ]{mewall@lanl.gov}

\date{\today}

\begin{abstract}

Inherent structure theory is used to discover strong connections
between simple characteristics of protein structure and the energy
landscape of a G$\bar{\bf \sf o}$ model. The potential energies and
vibrational free energies of inherent structures are highly
correlated, and both reflect simple measures of networks of native
contacts. These connections have important consequences for models of
protein dynamics and thermodynamics.

\end{abstract}


\maketitle


Protein activity is controlled by dynamical transitions among
conformational substates \cite{Frauenfelder85}; the transitions
may be understood in terms of motions on an energy landscape
\cite{Frauenfelder91}. Substates correspond to local minima in the
energy landscape, and transitions correspond to the hurdling of
barriers between minima. Interestingly, the protein energy landscape
resembles that of glasses \cite{Ansari85}.

Spin-glass models have yielded insight into properties of protein
energy landscapes \cite{Bryngelson87,Stein85} and the kinetics of
protein folding \cite{Bryngelson89}. The main motivation for using
spin-glass models rather than structural-glass models is that
spin-glass models are more analytically tractable; however, it has
long been recognized that structural-glass models might be
better-suited to describe proteins \cite{Stein85}. Indeed, protein
unfolding has been characterized as a rigidity transformation that is
similar to those seen in network glasses \cite{Rader02}.

Structural-glass-forming liquids have been fruitfully characterized
using inherent structure (IS) theory \cite{Stillinger82,Stillinger84},
which treats the energy landscape as a set of discrete basins that are
separated by saddles. Each basin contains a local minimum, called an
inherent structure, which is analogous to a protein conformational
substate. The dynamics are then naturally described as vibrations
about local minima, punctuated by transitions between neighboring
basins. A key assumption in IS theory is that vibrations are similar
about minima with the same potential energy; however, importantly, IS
theory allows for diversity among vibrations that have different
potential energies.

Guo and Thirumalai \cite{Guo96} have used IS theory to analyze
fluctuations in the neighborhood of the native state of a
coarse-grained model of a designed four-helix bundle
protein. Baumketner, Shea, and Hiwatari \cite{Baumketner03} have
applied IS theory to study the glass transition in a coarse-grained
model of a 16-residue polypeptide; by IS analysis of molecular
dynamics trajectories, they demonstrated the ability to rigorously
calculate the glass transition temperature due to freezing in the
native-state basin. In a more recent study, Nakagawa \& Peyrard
\cite{Nakagawa06} used IS theory to analyze the energy landscape of a
protein G B1 domain using a coarse-grained model, finding that the
density of minima increases exponentially with the
energy. Importantly, their analysis relied on an assumption that
vibrations are the same about all potential energy minima. However,
because proteins become less rigid as noncovalent bonds are broken
\cite{Rader02}, vibrations are expected to be different for different
minima, especially for minima with different potential
energies. Diversity in vibrations not only would change the density of
minima, but also would have important implications for the kinetics of
transitions among conformational substates \cite{Frauenfelder85};
however, if vibrations are the same for different minima, their role
in determining the kinetics of transitions would be trivial.

To characterize the diversity in vibrations among different protein
inherent structures, we used IS theory to analyze a coarse-grained
G$\bar{\rm o}$ model of the same protein fragment considered by
Nakagawa \& Peyrard \cite{Nakagawa06}: GB1, a protein G B1 domain
(Protein Data Bank \cite{Berman00} entry 2GB1
\cite{Gronenborn91}). GB1 has 56 amino acids and consists of a
four-stranded $\beta$-sheet packed against a single helix.

A configuration $x$ is represented by the set of ${\rm C}_\alpha$
positions, and the G$\bar{\rm o}$ model potential energy $U(x)$ for
${\rm C}_\alpha$ configurations $x$ of all proteins is similar to that
used by the GB1 studies in Refs.~\cite{Karanicolas02} and
\cite{Nakagawa06}:
\begin{eqnarray}
U(x)&=&\sum_{i=1}^{N-1}{K_b\over 2}\left(r_i-r_{i,0}\right)^2 + \sum_{i=1}^{N-2}{K_\theta\over 2}\left(\theta_i-\theta_{i,0}\right)^2\nonumber\\
& &+\sum_{i=1}^{N-3}{K_\phi\over 2}\left[1-\cos\left(2\phi_i-{\pi\over 2}\right)\right]\nonumber\\
& & +\sum_{i>j-3}^{NC}\epsilon \left[5\left({r_{ij,0}\over r_{ij}}\right)^{12}-6\left({r_{ij,0}\over r_{ij}}\right)^{10}\right]\nonumber\\
& &+\sum_{i>j-3}^{NNC}\epsilon\left({C\over r_{ij}}\right)^{12}.
\label{eq:U}
\end{eqnarray}
The first term in Eq.~(\ref{eq:U}) is the contribution from
neighboring backbone C$_\alpha$ bond distances, the second is from
angles between neighboring bonds, the third is from dihedral angles,
the fourth is from noncovalent interactions between native contacts,
and the fifth is from noncovalent interations between other pairs of
atoms. The crystal structure was used as the reference structure to
determine $r_{i,0}$, $\theta_{i,0}$, and $r_{ij,0}$, with native
contacts determined using a cutoff distance of $5.5~{\rm \AA}$. Other
parameter values are $K_b=200\epsilon_0~{\rm \AA}^{-2}$,
$K_\theta=40\epsilon_0~{\rm rad}^{-2}$, $K_\phi=0.3\epsilon_0$,
$\epsilon=0.18\epsilon_0$, and $C=4~{\rm \AA}$. The absolute energy
unit $\epsilon_0=1.89~{\rm Kcal}{\rm mol}^{-1}$ is determined as in
Ref.~\cite{Karanicolas02}, assuming a folding temperature of
350~K. Frustration is introduced through the dihedral angle terms,
which are not defined with respect to the reference structure.

Langevin dynamics simulations were performed using a time step
$0.0007\tau$ and a friction coefficient of $0.2/\tau$, where
$\tau=1.47$~ps (following Ref.~\cite{Karanicolas02}). The collapse
temperature $T_\theta$, at which extended and collapsed configurations
are approximately equally likely, was located by analysis of the
specific heat \cite{Nakagawa06}. A single trajectory at temperature
$T_\theta$ with $3\times 10^8$ time steps was sampled every $10^4$
steps to obtain an ensemble of $3\times 10^4$ inherent structures for
analysis. Local minima $e_\alpha$, corresponding to inherent
structures $\alpha$, were found using conjugate gradient minimization
terminated when a step resulted in an energy change of just
$10^{-12}$. The protein exhibited multiple transitions between
extended and collapsed states during the course of the simulation, and
the inherent structure ensembles exhibited a bimodal probability
distribution $P_{IS}(e_\alpha,T_\theta)$ of collapsed and extended
inherent-structure potential energies $e_\alpha$, similar to the
distribution in Ref.~\cite{Nakagawa06}.

Like a previous application of IS theory to proteins by Baumketner,
Shea, and Hiwatari \cite{Baumketner03}, we replace the configurational
integral in the partition function for an isolated protein with a sum
over contributions from individual inherent structures:
\begin{eqnarray}
Z&=&\left(\prod_{i=1}^{3N}\Lambda_i^{-1}\right)\frac{1}{\sigma}\int_V
dx\,e^{-U(x)/k_B T} \nonumber \\ &=& \sum_\alpha e^{-\beta
e_\alpha}\left(\prod_{i=1}^{3N}\Lambda_i^{-1}\right)\frac{1}{\sigma}\int_{R(\alpha)}
dx\,e^{-\left[U(x)-e_\alpha\right]/k_B T} \nonumber \\ &=&\sum_\alpha
e^{-\left[e_\alpha+F_v(\alpha,T)\right]/k_B T},
\label{eq:Z}
\end{eqnarray}
which defines the vibrational free energy $F_v(\alpha,T)$.
In Eq.~(\ref{eq:Z}), $R(\alpha)$ is the basin surrounding inherent
structure $\alpha$, $\Lambda_i$ is the thermal wavelength of atom $i$,
and $\sigma$ is a factor to account for symmetries.

\begin{figure}[t]
\begin{center}
\includegraphics[width=3.0in]{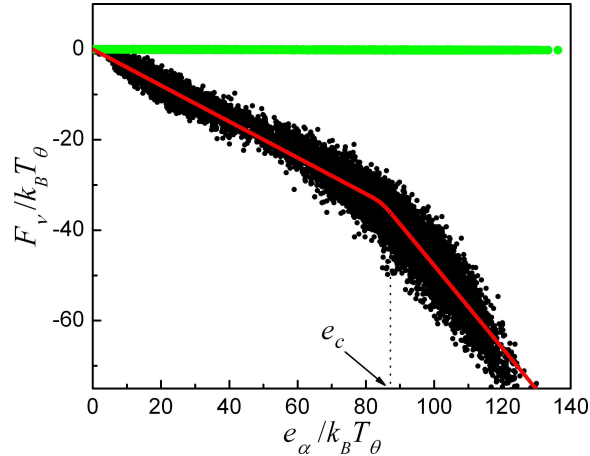}
\caption{Vibrational free energies $F_v$ of inherent structures
vs. their potential energies $e_\alpha$ (black points). The dependence
is well-modeled by Eq.~(\ref{eq:FitFv}) with $e_c = 88.4 k_B T_\theta$
(piecewise-linear red line). The contribution from the highest 1/3 of
the eigenvalues is constant (green points).}
\label{fig:Fv}
\end{center}
\end{figure}

Values of $F_v(\alpha,T)$, calculated as differences with respect to
the native structure $\alpha=0$ (the same holds for values of
$e_\alpha$), were estimated at the collapse temperature $T_\theta$
using a harmonic approximation,
\begin{equation}
F_v(\alpha,T)={k_B T\over 2} \sum_{i=7}^{3N}\ln{\lambda_i^{(\alpha)}\over\lambda_i^{(0)}},
\label{eq:FvHarm}
\end{equation}
where $\lambda_i^{(\alpha)}$ is the $i^{\rm th}$ eigenvalue of the
Hessian $h_{jk}=\partial^2U/\partial x_j \partial x_k$ calculated at
the energy minimum corresponding to inherent structure $\alpha$, and
$\lambda_i^{(0)}$ is the same for the ground-state inherent
structure. The sum is over all modes with nonzero frequency: we
neglect a contribution due to changes in the rotational entropy for
different inherent structures. Values of $F_v$ are similar for
inherent structures with a similar potential energy $e_\alpha$
(Fig.~\ref{fig:Fv}). The contribution to $F_v$ from the highest 1/3 of
the eigenvalues does not change for different inherent
structures. Interestingly, there is a gap in the eigenvalue spectrum
between the lowest 2/3 and the highest 1/3 of the eigenvalues; in
addition, only the highest 1/3 of the eigenvalues change when the
bond-distance force constant $K_b$ is increased by a factor of ten,
indicating that the corresponding modes describe the bond
vibrations. Therefore bond vibrations do not change significantly
among different inherent structures. However, the total $F_v$, which
includes contributions from the lowest 2/3 of the eigenvalues, changes
significantly with $e_\alpha$ (Fig.~\ref{fig:Fv}). The assumption of
constant $F_v$ by Nakagawa \& Peyrard \cite{Nakagawa06} therefore is
only valid for the modes that involve bond vibrations. This result is
consistent with studies of the loss of protein rigidity upon protein
unfolding \cite{Rader02}, and is also consistent with molecular
dynamics studies suggesting that vibrations can be diverse for
different protein conformational substates
\cite{Janezic95,vanVlijmen99}.

As demonstrated by the fit in Fig.~\ref{fig:Fv}, $F_v$ is well-modeled
using the function
\begin{eqnarray}
F_v(e_\alpha)=k_2e_\alpha&+&(k_2-k_1)k_B T_\theta \nonumber\\
& &\times\ln\left(e^{-e_c/k_B T_\theta}+e^{-e_\alpha/k_B T_\theta}\right).
\label{eq:FitFv}
\end{eqnarray}
Equation~(\ref{eq:FitFv}) is essentially a piecewise-linear function
with slope $k_1$ for $e_\alpha < e_c$, and slope $k_2$ for $e_\alpha >
e_c$. For GB1, $k_1=-0.40$, $k_2=-0.91$, and $e_c=88.4 k_B T_\theta$.

Inherent structure theory \cite{Stillinger82,Stillinger84} assumes
that $F_v(\alpha,T)=F_v(e_\alpha,T)$ (validated for the present
application in Fig.~\ref{fig:Fv}), and relates the vibrational free
energy $F_v(e_\alpha)$ and the probability distribution
$P_{IS}(e_\alpha)$ to the density of states $\Omega_{IS}(e_\alpha)$
through
\begin{equation}
P_{IS}(e_\alpha,T)={1\over Z}\Omega_{IS}(e_\alpha)e^{-e_\alpha/k_B T}e^{-F_v(e_\alpha,T)/k_B T}.
\label{eq:PIS}
\end{equation}
Given $\Omega_{IS}(e_0)=1$, $F_v(e_0,T)=0$, and $e_0=0$,
$\Omega_{IS}(e_\alpha)$ is given by
\begin{equation}
\Omega_{IS}(e_\alpha)=e^{e_\alpha/k_B T}e^{F_v(e_\alpha,T)/k_B T}{P_{IS}(e_\alpha,T)\over P_{IS}(e_0,T)},
\label{eq:OmegaIS}
\end{equation}
which generalizes a similar equation in Nakagawa \& Peyrard
\cite{Nakagawa06} to values of $F_v(e_\alpha,T)$ that vary with
$e_\alpha$. 

\begin{figure}[t]
\begin{center}
\includegraphics[width=3.0in]{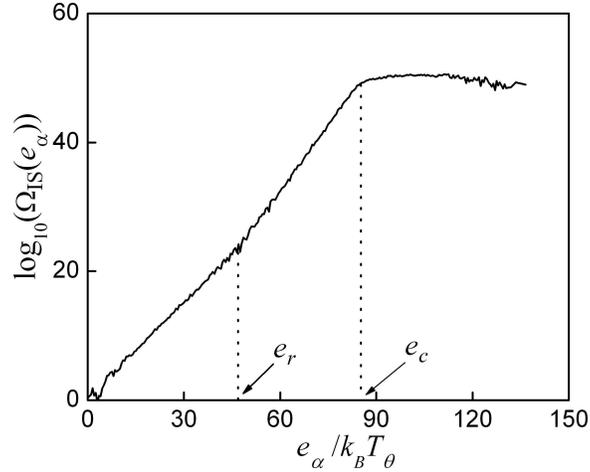}
\caption{Density of inherent structures
$\Omega_{IS}\left(e_\alpha\right)$. The knee at $e_r=47.4 k_B
T_\theta$ is due to a change in stress, and the plateau beginning at
roughly $e_c=88.4 k_B T_\theta$ is due to a change in rigidity; both are
understood in terms of the network of native contacts
(Figs.~\ref{fig:ealpha}, \ref{fig:n0}).}
\label{fig:Density}
\end{center}
\end{figure}

We used Eq.~(\ref{eq:OmegaIS}) along with the calculated
$P_{IS}(e_\alpha)$ and $F_v(e_\alpha)$ from Eq.~(\ref{eq:FitFv}) to
model the density of inherent structures $\Omega_{IS}(e_\alpha)$. At
energies below $e_c$, $\Omega_{IS}(e_\alpha)$ exhibits an exponential
increase, but with a slight increase in the exponent factor above an
energy $e_r$, giving rise to a knee in the plot of $\log
\Omega_{IS}(e_\alpha)$ vs. $e_\alpha$ (Fig.~\ref{fig:Density}). The
knee is located at a minimum in $P_{IS}(e_\alpha)$ between the
extended and collapsed states, and is thus associated with the
transition state. Such a knee was also seen in a previous model of
$\Omega_{IS}(e_\alpha)$ that did not consider vibrations
\cite{Nakagawa06}. Above $e_c$, $\Omega_{IS}(e_\alpha)$ plateaus and
decreases at the highest energies, which is a consequence of the
structure in both $P_{IS}(e_\alpha)$ and
$F_v(e_\alpha,T_\theta)$. Rather than being exponential in form
\cite{Nakagawa06}, from Eqs.~[\ref{eq:FitFv}] and [\ref{eq:OmegaIS}],
$\Omega_{IS}(e_\alpha)$ in this region has the form
\begin{equation}
\Omega_{IS}(e_\alpha)=e^{(1+k_2)e_\alpha/k_B T}e^{(k_1-k_2)e_c/k_B T_\theta}{P_{IS}(e_\alpha,T)\over P_{IS}(e_0,T)}.
\label{OmegaHigh}
\end{equation}
Because $k_2$ is close to -1 at $T_\theta$, the structure of
$\Omega_{IS}(e_\alpha)$ closely resembles that of
$P(e_\alpha,T_\theta)$ above $e_c$.

\begin{figure}[t]
\begin{center}
\includegraphics[width=3.0in]{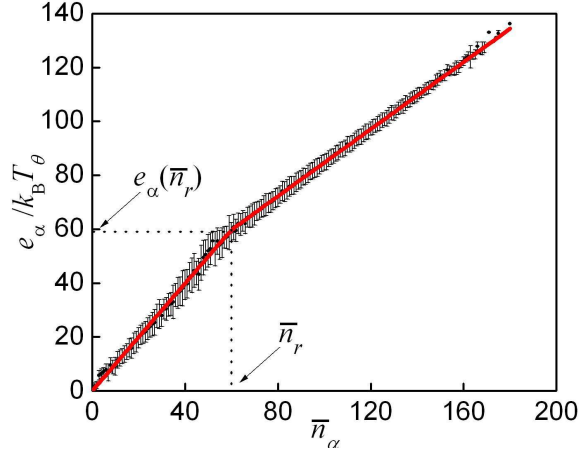}
\caption{Potential energy $e_\alpha$ vs. number of broken contacts
$\bar{n}_\alpha$. The dependence is well-modeled by
Eq.~(\ref{eq:ealpha}) (red line). The energy required to break a
native contact is approximately equal to the binding energy
$\epsilon$, with differences due to stress in the structure. The knee
corresponds to a change in stress at $\bar{n}_r=60$, where
$e_\alpha(\bar{n}_r)=59.6 k_B T_\theta$.}
\label{fig:ealpha}
\end{center}
\end{figure}

We found (Fig.~\ref{fig:ealpha}) that $e_\alpha$ is closely related to
the number of broken native contacts, $\bar{n}_\alpha$ through the
piecewise-linear function
\begin{equation}
e_\alpha=h_2\bar{n}_\alpha+(h_2-h_1)\ln\left(e^{-\bar{n}_r}+e^{-\bar{n}_\alpha}\right).
\label{eq:ealpha}
\end{equation}
The slopes $h_1$ and $h_2$ correspond to the amount of energy required
to break a native contact below ($h_1$) and above ($h_2$) a critical
number of broken contacts $\bar{n}_r$. Data for GB1 are well-modeled
using $h_1=0.997$, $h_2=0.622$, and $\bar{n}_r=60$. Below $\bar{n}_r$,
breaking a native contact requires more potential energy than above
$\bar{n}_r$. Therefore, $\bar{n}_r$ is associated with a change in
protein stress.

There are interesting connections between the structure of
$\Omega_{IS}(e_\alpha)$ below $e_c$ (Fig.~\ref{fig:Density}) and the
dependence of $e_\alpha$ on $\bar{n}_\alpha$
(Fig.~\ref{fig:ealpha}). The change in the slope of $\Omega_{IS}$ at
$e_r$ is closely related to the change in the slope of
$e_\alpha(\bar{n}_\alpha)$ at $\bar{n}_r$, suggesting that
$\Omega_{IS}$ has a simple exponential dependence on $\bar{n}_r$ below
$e_c$. However, the knee in $\Omega_{IS}$ occurs at $e_r = 47.4 k_B
T_\theta$, which is smaller than the value $e_\alpha(\bar{n}_r) = 59.6
k_B T_\theta$ at the knee in Fig.~\ref{fig:ealpha}. While the density
of inherent structures might truly be enhanced in the gap between
$e_r$ and $e_\alpha(\bar{n}_r)$, we note that the shift of $e_r$ with
respect to $e_\alpha(\bar{n}_r)$ might indicate that the inherent
structure basins associated with the transition state are especially
large (as noted above, $e_r$ is associated with the transition state),
and that the harmonic approximation might be especially ill-suited to
estimating their free energies for use in Eq.~(\ref{eq:OmegaIS}).

\begin{figure}[t]
\begin{center}
\includegraphics[width=3.0in]{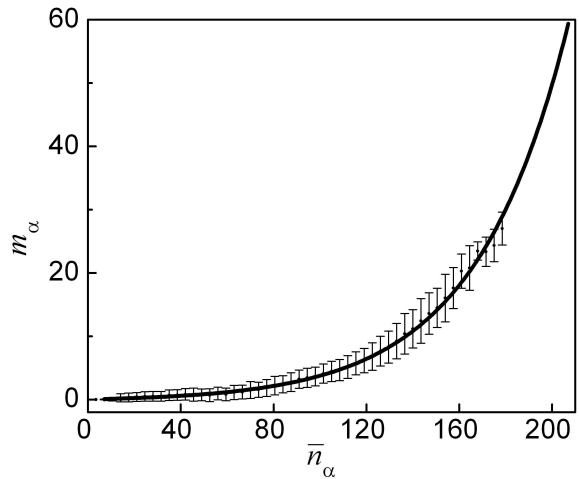}
\caption{Number of residues $m_\alpha$ for which all native contacts
are broken vs. the number of broken contacts $\bar{n}_\alpha$. The
dependence is well-modeled by Eq.~(\ref{eq:malpha}) (black line). Note
that $m_\alpha=55$ (close to the expected value of 56) at
$\bar{n}_\alpha=207$.}
\label{fig:n0}
\end{center}
\end{figure}

The source of the plateau in $\Omega_{IS}(e_\alpha)$ above $e_c$ may
be understood in terms of the dependence of the free energy on both
$\bar{n}_\alpha$ and the number of residues $m_\alpha$ for which all
native contacts are broken. As shown in Fig.~\ref{fig:n0}, a plot of
$m_\alpha$ vs. $\bar{n}_\alpha$ is well-modeled by the
function
\begin{equation}
m_\alpha = \frac{1}{3}\left(e^{\bar{n}_\alpha/40}-1\right),
\label{eq:malpha}
\end{equation}
supporting an expectation that breaking a contact is only likely to
create a residue with no native contacts at high $\bar{n}_\alpha$.
The following simple model for $F_v$ then successfully captures the
structure of $F_v$ in Fig.~\ref{fig:Fv}:
\begin{equation}
F_v(\bar{n}_\alpha) = \nu \bar{n}_\alpha + \mu m_\alpha,
\label{eq:Fvfit}
\end{equation}
with $m_\alpha$ given by Eq.~(\ref{eq:Fvfit}). Using $\nu=-0.32$ and
$\mu=-1.07$ yields good agreement between values of $F_v$ obtained
either directly from the Hessian or using Eq.~(\ref{eq:Fvfit}), with a
correlation coefficient of 0.993 for values calculated from all
inherent structures. We conclude that the change in the slope of $F_v$
vs. $e_\alpha$ at $e_c$, and therefore the plateau in
$\Omega_{IS}(e_\alpha)$ above $e_c$, is associated with an increase in
the likelihood that breaking a native contact will increase the number
of residues with no native contacts.

We found that protein stress and rigidity are closely tied to the
network of native contacts through Eqs.~[\ref{eq:ealpha}]
and~[\ref{eq:Fvfit}]. This finding is remeniscent of an analysis of
protein folding by Rader {\em et al.}~\cite{Rader02}, in which the
loss of network rigidity was associated with protein unfolding. It is
therefore tempting to associate the region between $e_r$ and $e_c$ in
Fig.~\ref{fig:Density} with the region of the mean coordination number
$\left<r\right>$ where Rader {\em et al.} found that proteins become
floppy and unfold. However, the present approach differs from that
used by Rader {\em et al.} in two key ways. First, because all residue
interactions in the present study are lumped into ${\rm C}_\alpha$
atoms, the coordination numbers are higher, and the relation of
coordination numbers to protein rigidity might be different than for
the all-atom models considered by Rader {\em et al.}  Second, whereas
the present results were obtained using a dynamical model, those
obtained by Rader {\em et al.}  were obtained using a static model of
the protein. It will be interesting to further explore connections
between the analyses based on IS theory and network rigidity; at
present, they provide complementary perspectives on the relationship
between protein dynamics and protein stress and rigidity.

The maximum in the density of states above $e_c$ is a consequence of
considering diversity in vibrations, and is not observed when uniform
vibrations are assumed \cite{Nakagawa06}. Interestingly, a similar
structure for the density of states, in which an exponential increase
is followed by a maximum, has been observed for many
structural-glass-forming liquids \cite{Stillinger84}. It will
therefore be interesting to improve the estimation of the density of
states by obtaining more accurate estimates of $F_v$ than are possible
using a harmonic approximation \cite{Baumketner03}.

Studies of two other G$\bar{\rm o}$ models of proteins yielded results
that are similar to those found here for GB1 (unpublished results),
suggesting the possibility that a simple phenomenological relationship
between the network of native contacts and the energy landscape might
exist for all G$\bar{\rm o}$ models.  It will be interesting to
explore this relationship for a large number of proteins and seek
representations in which it is identical for different proteins.
Discovery of such ``universality'' would enable the prediction of
important properties of the energy landscapes of G$\bar{\rm o}$ models
without performing numerical simulations.

It will be important to extend the present results to models whose
energy landscapes exhibit more frustration than G$\bar{\rm o}$ models.
For example, consider a modified model in which there is a weak
attractive interaction for non-native contacts. In contrast to the
simple relation illustrated in Fig.~\ref{fig:ealpha}, in such a model,
inherent structures with the same potential energy would likely have
diverse numbers of native contacts. However, by extending the
parameter space, the energy still might be simply related to a
combination of both the number of native contacts and the number of
non-native contacts. Similarly, the vibrational free energies might
exhibit diversity among inherent structures with the same energy, but
might still be simply related to both the number of native contacts
and non-native contacts through an equation analogous to
Eq.~(\ref{eq:Fvfit}). Ultimately, it will be interesting to
incrementally increase the complexity of the model, extending the
present results (as far as computationally feasible) to realistic,
all-atom models of proteins that include explicit solvent and other
effects that are important in controlling protein function. Use of
such all-atom models will enable the link between analyses based on IS
theory and network rigidity to be further explored.

The present results demonstrate that simple connections to protein
structure are hidden within the energy landscape of a G$\bar{\rm o}$
model. The potential energies and vibrational free energies of
inherent structures are highly correlated, and both reflect simple
measures of networks of native contacts. Through use of IS theory,
these regularities should enable significant simplification of
thermodynamic models of proteins
\cite{Stillinger82,Stillinger84,Nakagawa06}.

\begin{acknowledgments}
We gratefully acknowledge Arthur Voter and Donald Jacobs for
discussions. This work was supported by the Department of Energy.
\end{acknowledgments}
\bibliography{Vib}

\begin{thebibliography}{17}
\expandafter\ifx\csname natexlab\endcsname\relax\def\natexlab#1{#1}\fi
\expandafter\ifx\csname bibnamefont\endcsname\relax
  \def\bibnamefont#1{#1}\fi
\expandafter\ifx\csname bibfnamefont\endcsname\relax
  \def\bibfnamefont#1{#1}\fi
\expandafter\ifx\csname citenamefont\endcsname\relax
  \def\citenamefont#1{#1}\fi
\expandafter\ifx\csname url\endcsname\relax
  \def\url#1{\texttt{#1}}\fi
\expandafter\ifx\csname urlprefix\endcsname\relax\def\urlprefix{URL }\fi
\providecommand{\bibinfo}[2]{#2}
\providecommand{\eprint}[2][]{\url{#2}}

\bibitem[{\citenamefont{Frauenfelder and Wolynes}(1985)}]{Frauenfelder85}
\bibinfo{author}{\bibfnamefont{H.}~\bibnamefont{Frauenfelder}}
  \bibnamefont{and} \bibinfo{author}{\bibfnamefont{P.~G.}
  \bibnamefont{Wolynes}}, \bibinfo{journal}{Science}
  \textbf{\bibinfo{volume}{229}}, \bibinfo{pages}{337} (\bibinfo{year}{1985}).

\bibitem[{\citenamefont{Frauenfelder et~al.}(1991)\citenamefont{Frauenfelder,
  Sligar, and Wolynes}}]{Frauenfelder91}
\bibinfo{author}{\bibfnamefont{H.}~\bibnamefont{Frauenfelder}},
  \bibinfo{author}{\bibfnamefont{S.~G.} \bibnamefont{Sligar}},
  \bibnamefont{and} \bibinfo{author}{\bibfnamefont{P.~G.}
  \bibnamefont{Wolynes}}, \bibinfo{journal}{Science}
  \textbf{\bibinfo{volume}{254}}, \bibinfo{pages}{1598} (\bibinfo{year}{1991}).

\bibitem[{\citenamefont{Ansari et~al.}(1985)\citenamefont{Ansari, Berendzen,
  Bowne, Frauenfelder, Iben, Sauke, Shyamsunder, and Young}}]{Ansari85}
\bibinfo{author}{\bibfnamefont{A.}~\bibnamefont{Ansari}},
  \bibinfo{author}{\bibfnamefont{J.}~\bibnamefont{Berendzen}},
  \bibinfo{author}{\bibfnamefont{S.~F.} \bibnamefont{Bowne}},
  \bibinfo{author}{\bibfnamefont{H.}~\bibnamefont{Frauenfelder}},
  \bibinfo{author}{\bibfnamefont{I.~E.} \bibnamefont{Iben}},
  \bibinfo{author}{\bibfnamefont{T.~B.} \bibnamefont{Sauke}},
  \bibinfo{author}{\bibfnamefont{E.}~\bibnamefont{Shyamsunder}},
  \bibnamefont{and} \bibinfo{author}{\bibfnamefont{R.~D.} \bibnamefont{Young}},
  \bibinfo{journal}{Proc Natl Acad Sci U S A} \textbf{\bibinfo{volume}{82}},
  \bibinfo{pages}{5000} (\bibinfo{year}{1985}).

\bibitem[{\citenamefont{Bryngelson and Wolynes}(1987)}]{Bryngelson87}
\bibinfo{author}{\bibfnamefont{J.~D.} \bibnamefont{Bryngelson}}
  \bibnamefont{and} \bibinfo{author}{\bibfnamefont{P.~G.}
  \bibnamefont{Wolynes}}, \bibinfo{journal}{Proc Natl Acad Sci U S A}
  \textbf{\bibinfo{volume}{84}}, \bibinfo{pages}{7524} (\bibinfo{year}{1987}).

\bibitem[{\citenamefont{Stein}(1985)}]{Stein85}
\bibinfo{author}{\bibfnamefont{D.~L.} \bibnamefont{Stein}},
  \bibinfo{journal}{Proc Natl Acad Sci U S A} \textbf{\bibinfo{volume}{82}},
  \bibinfo{pages}{3670} (\bibinfo{year}{1985}).

\bibitem[{\citenamefont{Bryngelson and Wolynes}(1989)}]{Bryngelson89}
\bibinfo{author}{\bibfnamefont{J.~D.} \bibnamefont{Bryngelson}}
  \bibnamefont{and} \bibinfo{author}{\bibfnamefont{P.~G.}
  \bibnamefont{Wolynes}}, \bibinfo{journal}{J Phys Chem}
  \textbf{\bibinfo{volume}{93}}, \bibinfo{pages}{6902} (\bibinfo{year}{1989}).

\bibitem[{\citenamefont{Rader et~al.}(2002)\citenamefont{Rader, Hespenheide,
  Kuhn, and Thorpe}}]{Rader02}
\bibinfo{author}{\bibfnamefont{A.~J.} \bibnamefont{Rader}},
  \bibinfo{author}{\bibfnamefont{B.~M.} \bibnamefont{Hespenheide}},
  \bibinfo{author}{\bibfnamefont{L.~A.} \bibnamefont{Kuhn}}, \bibnamefont{and}
  \bibinfo{author}{\bibfnamefont{M.~F.} \bibnamefont{Thorpe}},
  \bibinfo{journal}{Proc Natl Acad Sci U S A} \textbf{\bibinfo{volume}{99}},
  \bibinfo{pages}{3540} (\bibinfo{year}{2002}).

\bibitem[{\citenamefont{Stillinger and Weber}(1982)}]{Stillinger82}
\bibinfo{author}{\bibfnamefont{F.~H.} \bibnamefont{Stillinger}}
  \bibnamefont{and} \bibinfo{author}{\bibfnamefont{T.~A.} \bibnamefont{Weber}},
  \bibinfo{journal}{Phys Rev A} \textbf{\bibinfo{volume}{25}},
  \bibinfo{pages}{978} (\bibinfo{year}{1982}).

\bibitem[{\citenamefont{Stillinger and Weber}(1984)}]{Stillinger84}
\bibinfo{author}{\bibfnamefont{F.~H.} \bibnamefont{Stillinger}}
  \bibnamefont{and} \bibinfo{author}{\bibfnamefont{T.~A.} \bibnamefont{Weber}},
  \bibinfo{journal}{Science} \textbf{\bibinfo{volume}{225}},
  \bibinfo{pages}{983} (\bibinfo{year}{1984}).

\bibitem[{\citenamefont{Guo and Thirumalai}(1996)}]{Guo96}
\bibinfo{author}{\bibfnamefont{Z.}~\bibnamefont{Guo}} \bibnamefont{and}
  \bibinfo{author}{\bibfnamefont{D.}~\bibnamefont{Thirumalai}},
  \bibinfo{journal}{J Mol Biol} \textbf{\bibinfo{volume}{263}},
  \bibinfo{pages}{323} (\bibinfo{year}{1996}).

\bibitem[{\citenamefont{Baumketner et~al.}(2003)\citenamefont{Baumketner, Shea,
  and Hiwatari}}]{Baumketner03}
\bibinfo{author}{\bibfnamefont{A.}~\bibnamefont{Baumketner}},
  \bibinfo{author}{\bibfnamefont{J.-E.} \bibnamefont{Shea}}, \bibnamefont{and}
  \bibinfo{author}{\bibfnamefont{Y.}~\bibnamefont{Hiwatari}},
  \bibinfo{journal}{Phys Rev E} \textbf{\bibinfo{volume}{67}},
  \bibinfo{pages}{011912} (\bibinfo{year}{2003}).

\bibitem[{\citenamefont{Nakagawa and Peyrard}(2006)}]{Nakagawa06}
\bibinfo{author}{\bibfnamefont{N.}~\bibnamefont{Nakagawa}} \bibnamefont{and}
  \bibinfo{author}{\bibfnamefont{M.}~\bibnamefont{Peyrard}},
  \bibinfo{journal}{Proc Natl Acad Sci U S A} \textbf{\bibinfo{volume}{103}},
  \bibinfo{pages}{5279} (\bibinfo{year}{2006}).

\bibitem[{\citenamefont{Berman et~al.}(2000)\citenamefont{Berman, Westbrook,
  Feng, Gilliland, Bhat, Weissig, Shindyalov, and Bourne}}]{Berman00}
\bibinfo{author}{\bibfnamefont{H.~M.} \bibnamefont{Berman}},
  \bibinfo{author}{\bibfnamefont{J.}~\bibnamefont{Westbrook}},
  \bibinfo{author}{\bibfnamefont{Z.}~\bibnamefont{Feng}},
  \bibinfo{author}{\bibfnamefont{G.}~\bibnamefont{Gilliland}},
  \bibinfo{author}{\bibfnamefont{T.~N.} \bibnamefont{Bhat}},
  \bibinfo{author}{\bibfnamefont{H.}~\bibnamefont{Weissig}},
  \bibinfo{author}{\bibfnamefont{I.~N.} \bibnamefont{Shindyalov}},
  \bibnamefont{and} \bibinfo{author}{\bibfnamefont{P.~E.}
  \bibnamefont{Bourne}}, \bibinfo{journal}{Nucleic Acids Res}
  \textbf{\bibinfo{volume}{28}}, \bibinfo{pages}{235} (\bibinfo{year}{2000}).

\bibitem[{\citenamefont{Gronenborn et~al.}(1991)\citenamefont{Gronenborn,
  Filpula, Essig, Achari, Whitlow, Wingfield, and Clore}}]{Gronenborn91}
\bibinfo{author}{\bibfnamefont{A.~M.} \bibnamefont{Gronenborn}},
  \bibinfo{author}{\bibfnamefont{D.~R.} \bibnamefont{Filpula}},
  \bibinfo{author}{\bibfnamefont{N.~Z.} \bibnamefont{Essig}},
  \bibinfo{author}{\bibfnamefont{A.}~\bibnamefont{Achari}},
  \bibinfo{author}{\bibfnamefont{M.}~\bibnamefont{Whitlow}},
  \bibinfo{author}{\bibfnamefont{P.~T.} \bibnamefont{Wingfield}},
  \bibnamefont{and} \bibinfo{author}{\bibfnamefont{G.~M.} \bibnamefont{Clore}},
  \bibinfo{journal}{Science} \textbf{\bibinfo{volume}{253}},
  \bibinfo{pages}{657} (\bibinfo{year}{1991}).

\bibitem[{\citenamefont{Karanicolas and Brooks}(2002)}]{Karanicolas02}
\bibinfo{author}{\bibfnamefont{J.}~\bibnamefont{Karanicolas}} \bibnamefont{and}
  \bibinfo{author}{\bibfnamefont{C.~L.} \bibnamefont{Brooks}},
  \bibinfo{journal}{Protein Sci} \textbf{\bibinfo{volume}{11}},
  \bibinfo{pages}{2351} (\bibinfo{year}{2002}).

\bibitem[{\citenamefont{Janezic et~al.}(1995)\citenamefont{Janezic, Venable,
  and Brooks}}]{Janezic95}
\bibinfo{author}{\bibfnamefont{D.}~\bibnamefont{Janezic}},
  \bibinfo{author}{\bibfnamefont{R.~M.} \bibnamefont{Venable}},
  \bibnamefont{and} \bibinfo{author}{\bibfnamefont{B.~R.}
  \bibnamefont{Brooks}}, \bibinfo{journal}{J Comput Chem}
  \textbf{\bibinfo{volume}{16}}, \bibinfo{pages}{1554} (\bibinfo{year}{1995}).

\bibitem[{\citenamefont{van Vlijmen and Karplus}(1999)}]{vanVlijmen99}
\bibinfo{author}{\bibfnamefont{H.~W.~T.} \bibnamefont{van Vlijmen}}
  \bibnamefont{and} \bibinfo{author}{\bibfnamefont{M.}~\bibnamefont{Karplus}},
  \bibinfo{journal}{J Phys Chem B} \textbf{\bibinfo{volume}{103}},
  \bibinfo{pages}{3009} (\bibinfo{year}{1999}).

\end{thebibliography}

\end{document}